\documentclass[aps,prl,reprint,superscriptaddress,floatfix]{revtex4-2}

\usepackage{graphicx}
\usepackage{amsmath}
\usepackage{amssymb}
\usepackage{bm}
\usepackage[T1]{fontenc}
\usepackage[utf8]{inputenc}

\newcommand{\Cgeo}{C_{\mathrm{geo}}}
\newcommand{\Qdispl}{Q_{\mathrm{displ}}}

\newcommand{\Vbi}{V_{\mathrm{bi}}}
\newcommand{\Vbiz}{V_{\mathrm{bi},0}}
\newcommand{\Ebulk}{E_{\mathrm{bulk}}}
\newcommand{\an}{\mathrm{an}}
\newcommand{\cat}{\mathrm{cat}}
\newcommand{\mim}{\textit{m-i-m}}
\newcommand{\CV}{\textit{C-V}}

\begin{document}

\title{Capacitance of Undoped Thin-Film Diodes}

\author{Oskar J. Sandberg}
\email{oskar.sandberg@abo.fi}
\affiliation{Physics, Faculty of Science and Engineering, {\AA}bo Akademi University, Henrikinkatu 2, Turku 20500, Finland}

\author{Mathias Nyman}
\affiliation{Physics, Faculty of Science and Engineering, {\AA}bo Akademi University, Henrikinkatu 2, Turku 20500, Finland}

\author{Stefan Zeiske}
\affiliation{Department of Chemistry, Northwestern University, Evanston, Illinois 60208, United States}

\author{Staffan Dahlstr\"om}
\affiliation{Physics, Faculty of Science and Engineering, {\AA}bo Akademi University, Henrikinkatu 2, Turku 20500, Finland}

\author{Johannes Benduhn}
\affiliation{Dresden Integrated Center for Applied Physics and Photonic Materials (IAPP) and Institute of Applied Physics, Technische Universit\"at Dresden, N\"othnitzer Str. 61, 01187 Dresden, Germany}
\affiliation{Center for Technology Development, German Center for Astrophysics, Postplatz 1, 02826 G\"orlitz, Germany}

\author{Ardalan Armin}
\affiliation{Centre for Integrative Semiconductor Materials (CISM), Department of Physics, Swansea University, Swansea SA1 8EN, United Kingdom}

\date{\today}

\begin{abstract}
The capacitance of thin-film diodes based on undoped semiconductors are dominated by injected charge carriers rather than doping-induced carriers. However, an analytical framework of the capacitance in these devices has remained elusive. Here, we derive an analytical description of the capacitance, fully accounting for injected charge carrier and electrode charge effects. Based on these findings, a method to extract the built-in voltage in these devices is presented. The theoretical framework is substantiated numerically by drift-diffusion simulations and experimentally on organic solar cells.
\end{abstract}

\maketitle

Thin-film diode devices based on undoped organic semiconductors have shown great promise for a variety of emerging applications, including light-emitting diodes, solar cells, sensors and photodetectors~\cite{Forrest2000}. A key parameter of any thin-film diode is the built-in potential, providing the driving force for charge extraction in photovoltaic devices and controlling the onset for injection in light-emitting diodes. Capacitance-voltage (\CV) measurements have been widely used to probe the built-in voltage in semiconductor diode devices~\cite{Sze1981,Neukom2018}. To this end, the \CV{} characteristics are routinely analysed using Mott-Schottky analysis, where the built-in potential is extracted from the intercept in an inverse-square of the capacitance ($1/C^{2}$) \emph{vs} voltage plot~\cite{Sze1981}. However, Mott-Schottky analysis is only valid for sufficiently thick active layers with uniform doping. In contrast, Mott-Schottky analysis has been demonstrated to be highly unreliable in devices with thin active layers and/or low doping levels~\cite{Mingebach2011,Kirchartz2012,Nigam2013}. These devices instead behave as metal-insulator-metal (\mim) diodes, characterized by fully depleted, effectively undoped active layers~\cite{Kirchartz2015,Dahlstrom2019}.

The dark capacitance of a \mim{} diode device, at low voltages, is ideally described by its geometric capacitance~\cite{Shao1961},
\begin{equation}
\Cgeo = \frac{\varepsilon\varepsilon_{0}}{d},
\label{eq:1}
\end{equation}
where $\varepsilon$ is the relative permittivity of the active layer, $\varepsilon_{0}$ is the vacuum permittivity, and $d$ is the thickness of the active layer. However, in devices with injecting contacts (e.g., light-emitting diodes and solar cells), the capacitance typically exceeds $\Cgeo$, generally showing a pronounced voltage dependence~\cite{Shrotriya2005,vanMensfoort2008a,Kiermasch2018,Zeiske2022}. This has been attributed to injected charge carriers that have diffused into the active layer from the contacts~\cite{Kirchartz2012,vanMensfoort2008a}, resulting in an additional chemical contribution to the capacitance~\cite{Bisquert2003}. While analytical models for incorporating the chemical (or diffusion) capacitance have been proposed in the past~\cite{Tripathi2013,Nandal2018}, these descriptions do not account for space charge effects, influencing both the charge within the active layer and the charge at the electrodes, and thus the associated electrode capacitance.

Injected charge carriers are known to induce space charge effects and energy level bending near the contacts in thin-film devices~\cite{vanMensfoort2008b,Lange2011,deBruyn2013,Sandberg2024}. On the other hand, any space charge within the active layer will inevitably also influence the charge on the electrodes~\cite{Hawks2015}, giving rise to an electrode capacitance that depends non-trivially on the voltage~\cite{Zonno2019}. To the best of our knowledge, a rigorous analytical treatment which accounts for the interdependence between the space charge induced by injected carriers, their influence on the electrode charge, and the built-in potential has not yet been presented.

In this work, we present a theoretical framework incorporating the mutual dependence between injected charge carriers, the electrode charge, and the capacitance of sandwich-type devices based on undoped semiconductors. Based on these considerations, we derive analytical expressions describing the dark \CV{} characteristics of \mim{} diodes at low frequencies, and its relation to the built-in potential. The analytical framework is validated by numerical drift-diffusion simulations. Based on the analytical findings, a method to extract the built-in potential from \CV{} characteristics of thin-film diode devices is proposed and demonstrated experimentally on organic solar cell devices.

We consider a thin-film diode structure constituting an active semiconductor layer sandwiched between two electrodes: a hole-injecting anode and an electron-injecting cathode. The anode contact is situated at $x=0$ and the cathode contact at $x=d$, where $x$ denotes the position in the device. The active semiconductor layer is assumed to be undoped, corresponding to the case of a thin enough device where space charge effects induced by traps and dopants are negligible. In this case, the electron and hole densities, $n(x)$ and $p(x)$, in the active layer are related to the electric field $E(x)$ through the Poisson equation:
\begin{equation}
\frac{\partial E(x)}{\partial x} = \frac{\rho(x)}{\varepsilon\varepsilon_{0}},
\label{eq:2}
\end{equation}
where $\rho(x) = q\left[p(x)-n(x)\right]$ is the space charge density and $q$ is the elementary charge. Conversely, $E(x)$ is related to the applied voltage $V$ via
\begin{equation}
V - \Vbiz = \int_{0}^{d} E(x)\,dx,
\label{eq:3}
\end{equation}
where $\Vbiz = \left[\Phi_{\an}-\Phi_{\cat}\right]/q$ is the total built-in potential determined by the difference between the work function at the anode ($\Phi_{\an}$) and cathode ($\Phi_{\cat}$) contact. Finally, the electron [hole] density at the cathode [anode] is assumed to be given by $n_{\cat} = N_{c}\exp\left(-\varphi_{\cat}/kT\right)$ [$p_{\an} = N_{v}\exp\left(-\varphi_{\an}/kT\right)$], where $\varphi_{\cat} = \Phi_{\cat}-\chi_{p}$ [$\varphi_{\an} = \chi_{n}-\Phi_{\an}$] is the injection barrier at the cathode [anode], $\chi_{n}$ [$\chi_{p}$] is the electron affinity [ionization potential], $N_{c}$ [$N_{v}$] is the effective density of electron [hole] states, $k$ is the Boltzmann constant, and $T$ the temperature. Note that $q\Vbiz = E_{g}-\varphi_{\an}-\varphi_{\cat}$, where $E_{g} = \chi_{p}-\chi_{n}$ is the energy level gap.

\begin{figure*}[t]
\centering
\includegraphics[width=0.86\textwidth]{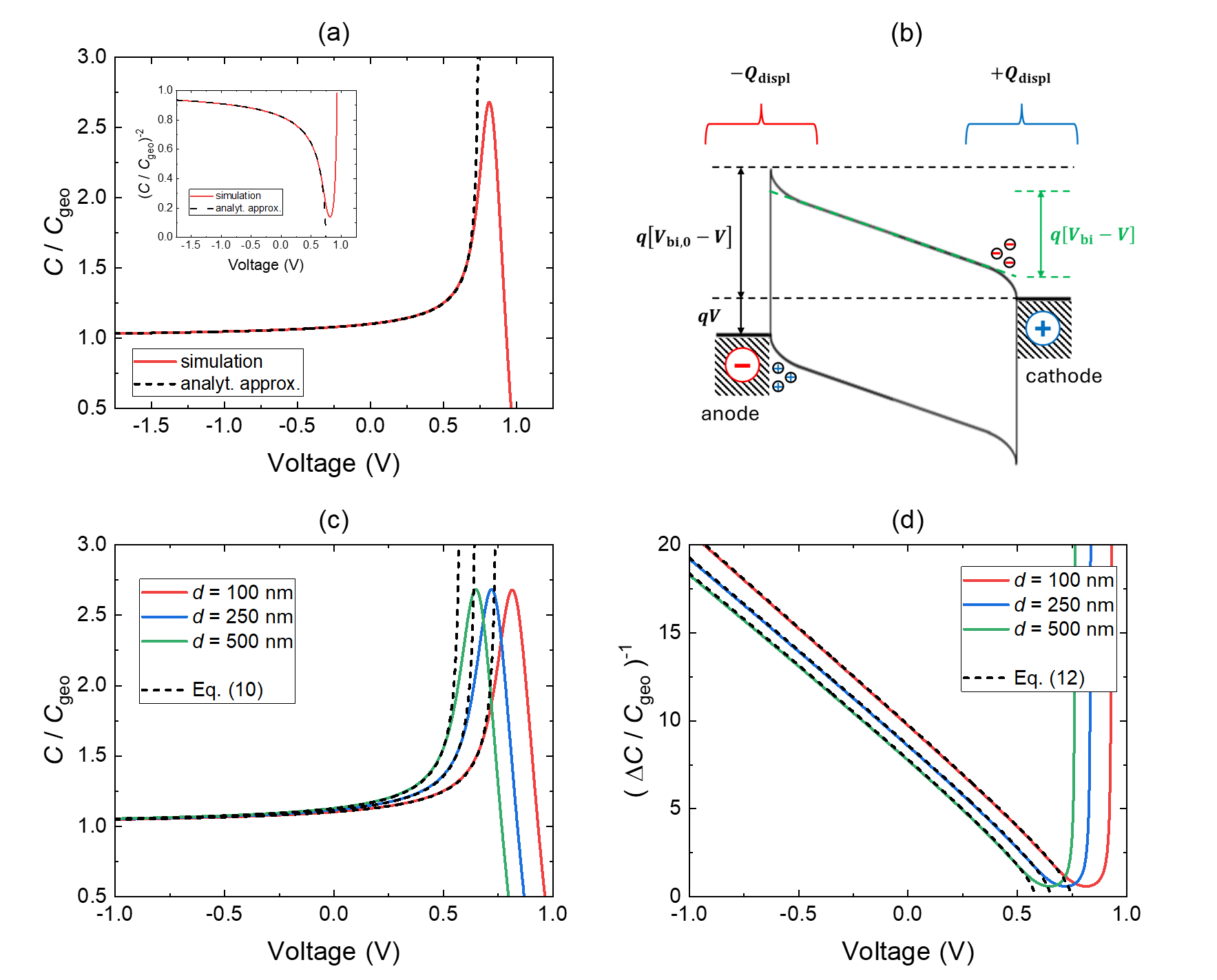}
\caption{(a) Simulated \CV{} characteristics of a 100~nm thick organic \mim{} diode with two ohmic contacts. The capacitance is normalized to the geometric capacitance $\Cgeo$. In the inset of (a) the corresponding Mott-Schottky plot (i.e., $C^{-2}$ vs $V$) is included for comparison. (b) The energy level diagram of an \mim{} diode for a forward biased voltage $V$ below the built-in potential. The charge $-\Qdispl$ at the anode side is balanced by an equal but opposite charge $\Qdispl$ at the cathode side. (c) Normalized \CV{} characteristics of a \mim{} diode with two ohmic contacts simulated at different active layer thicknesses. (d) The same curves from (c) but replotted as $\left(\Delta C/\Cgeo\right)^{-1}$ vs $V$. The simulated curves are depicted by solid lines, while the analytical expressions Eq.~(\ref{eq:10}) and Eq.~(\ref{eq:12}) are indicated by dashed lines.}
\label{fig:1}
\end{figure*}

The simulated \CV{} characteristics of an exemplary organic \mim{} diode device with ohmic contacts and $d=100$~nm is shown in Fig.~\ref{fig:1}(a). For the simulations, a numerical drift-diffusion model is used~\cite{Sandberg2019}; the parameters used are provided in the Supplemental Material~\cite{SupplementalMaterial}. The capacitance is obtained at a sufficiently low frequency, in a regime where the \CV{} plot is not limited by transport and independent of frequency. A pronounced voltage dependence is observed, with the capacitance increasing with increasing forward bias. While the capacitance is slowly saturating towards the geometric capacitance in the reverse-bias. Importantly, as demonstrated in the inset of Fig.~\ref{fig:1}(a), the inverse square of the capacitance ($1/C^{2}$) does not follow the linear voltage dependence expected for a device dominated by (uniformly distributed) doping-induced carriers; the obtained $1/C^{2}$ vs $V$ plot instead suggests a strongly nonuniform charge carrier distribution, characteristic of injected carriers.

To obtain analytical insight into this behaviour, we consider steady-state conditions, corresponding to the low-frequency limit. Subsequently, an increase $\delta V$ in the applied voltage $V$ will give rise to a change in the electric field across the device. This will on one hand induce a corresponding change in the surface charge stored at the electrodes; on the other hand, modify the electron and hole density profiles within the active layer. In the process, a charge $\delta\Qdispl$ will be displaced across the system. The associated device capacitance is given by
\begin{equation}
C = \frac{\partial\Qdispl}{\partial V},
\label{eq:4}
\end{equation}
where
\begin{equation}
\Qdispl = \frac{Q_{\an}-Q_{\cat}}{2} + \frac{qd}{2}\left(\overline{n}+\overline{p}\right),
\label{eq:5}
\end{equation}
in accordance with the law of charge conservation~\cite{SupplementalMaterial}. Here, $Q_{\an} = \varepsilon\varepsilon_{0}E(0)$ and $Q_{\cat} = -\varepsilon\varepsilon_{0}E(d)$ are the surface charge densities at the anode and cathode, respectively, while $\overline{n} = (1/d)\int_{0}^{d} n(x)\,dx$ and $\overline{p} = (1/d)\int_{0}^{d} p(x)\,dx$.

Eq.~(\ref{eq:5}) represents the displaced charge in a \mim{} diode (with injecting contacts) for the case of an undoped, trap free active layer [see Fig.~\ref{fig:1}(b)]. The first term accounts for the charge stored at the electrodes~\cite{Hawks2015}, while the second term accounts for a chemical charge stored within the active layer~\cite{Hartnagel2023}. Note that in the absence of injected carriers ($n=p=0$), the electric field will be uniform and given by $E = \left[V-\Vbiz\right]/d$; in this limit, $\Qdispl = \Cgeo\left[V-\Vbiz\right]$ and $C=\Cgeo$. In the presence of injected carriers, however, additional voltage dependences in both the chemical charge within the active layer and the charge stored at the electrodes are generally obtained, giving rise to an enhanced $C$ relative to $\Cgeo$.

To account for injected carriers, we note that, for $V$ below $\Vbiz$, the charge within the active layer is primarily concentrated to regions near the electrodes. Holes dominate the space charge near the hole-injecting anode contact, while electrons dominate the space charge near the electron-injecting cathode contact. Subsequently, we may divide the active layer into a hole-dominated region ($0\leq x<x^{*}$) and an electron-dominated region ($x^{*}<x\leq d$), where $x^{*}$ is the width of the hole-dominated region. Inside the active layer sufficiently far away from both contacts, the space charge is negligible, and $E(x)$ is constant and given by the bulk field $\Ebulk$. Then, integrating Eq.~(\ref{eq:2}) in the hole-dominated region yields $q\overline{p}d = \varepsilon\varepsilon_{0}\left[\Ebulk-E(0)\right]$. A similar integration in the electron-dominated region reveals $q\overline{n}d = \varepsilon\varepsilon_{0}\left[\Ebulk-E(d)\right]$. Hence, Eq.~(\ref{eq:5}) simplifies to
\begin{equation}
\Qdispl = \varepsilon\varepsilon_{0}\Ebulk.
\label{eq:6}
\end{equation}
Note that for the case of a hole-only [electron-only] diode with an injecting anode [cathode] contact and a non-injecting cathode [anode] contact, we get $\Ebulk = E(d)$ [$\Ebulk = E(0)$].

To obtain a general expression for $\Ebulk$, we approximate the hole density for $x<x^{*}$ by $p(x) = p_{\an}\exp\left(-q\phi(x)/kT\right)$, where $\phi(x) = -\int_{0}^{x} E(x')\,dx'$. This approximation is valid as long as the hole quasi-Fermi level remains flat in the accumulation region near the anode. Similarly, the electron density for $x>x^{*}$ is approximated by $n(x) = n_{\cat}\exp\left(q\left[\phi(x)-\phi(d)\right]/kT\right)$. Then, by assuming that $\rho(x)\approx qp(x)$ for $x<x^{*}$, and $\rho(x)\approx -qn(x)$ for $x>x^{*}$, Eq.~(\ref{eq:2}) can be solved separately in these regions~\cite{Sandberg2024,deLevie1972a,deLevie1972b,Cheyns2008}. Assuming that $E(x)\rightarrow\Ebulk$ and $\rho(x)\rightarrow 0$ sufficiently far away from both contacts, we find
\begin{widetext}
\begin{equation}
E(x) =
\begin{cases}
\Ebulk\coth\left[\dfrac{q\left|\Ebulk\right|x}{2kT} + \sinh^{-1}\left(\dfrac{q\left|\Ebulk\right|\lambda_{\an}}{2kT}\right)\right], & \text{if } x<x^{*},\\[3ex]
\Ebulk\coth\left[\dfrac{q\left|\Ebulk\right|(d-x)}{2kT} + \sinh^{-1}\left(\dfrac{q\left|\Ebulk\right|\lambda_{\cat}}{2kT}\right)\right], & \text{if } x>x^{*},
\end{cases}
\label{eq:7}
\end{equation}
\end{widetext}
where $\lambda_{\an} = \sqrt{2\varepsilon\varepsilon_{0}kT/\left(q^{2}p_{\an}\right)}$ and $\lambda_{\cat} = \sqrt{2\varepsilon\varepsilon_{0}kT/\left(q^{2}n_{\cat}\right)}$ are the Debye screening lengths at the anode and cathode, respectively.

Accordingly, the field depends strongly on $x$ near the injecting contacts but saturates rapidly to the constant value $E(x)\rightarrow\Ebulk$ inside the active layer for $\left|\Ebulk\right|\gg 2kT/qd$. After applying the boundary condition Eq.~(\ref{eq:3}) to Eq.~(\ref{eq:7}), we obtain
\begin{equation}
\Ebulk = \frac{V-\Vbi(V)}{d},
\label{eq:8}
\end{equation}
where $\Vbi(V)$ represents the effective built-in potential given by
\begin{widetext}
\begin{equation}
\Vbi(V) = \Vbiz
- \frac{2kT}{q}\ln\left[\frac{1}{2}\left\{\sqrt{1+\left(\frac{2kTd}{q\left[\Vbi(V)-V\right]\lambda_{\an}}\right)^{2}}+1\right\}\right]
- \frac{2kT}{q}\ln\left[\frac{1}{2}\left\{\sqrt{1+\left(\frac{2kTd}{q\left[\Vbi(V)-V\right]\lambda_{\cat}}\right)^{2}}+1\right\}\right]
\label{eq:9}
\end{equation}
\end{widetext}
and depends on the applied voltage. Note that $\Vbi(V)$ is generally reduced compared to $\Vbiz$, owing to the energy-level bending induced by accumulation of injected charge carriers near the contacts, reflecting the true built-in electric field experienced by carriers inside the active layer. The corresponding magnitude of the energy-level bending at the anode and cathode contact is quantified by the second and third term on the right-hand-side of Eq.~(\ref{eq:9}), respectively. A schematic energy level diagram of the situation is shown in Fig.~\ref{fig:1}(b).

Finally, after inserting Eq.~(\ref{eq:8}) and (\ref{eq:9}) into Eq.~(\ref{eq:6}) and making use of implicit differentiation, the capacitance of an \mim{} diode is obtained as
\begin{equation}
C(V) = \Cgeo\times\left[1-\frac{2kT}{q\left[\Vbi(V)-V\right]}\times\eta(V)\right]^{-1}
\label{eq:10}
\end{equation}
where $\eta = \eta_{\an}+\eta_{\cat}$ with
\begin{equation}
\eta_{\an(\cat)}(V) = 1-\frac{1}{\sqrt{1+\left(\dfrac{2kTd}{q\left[\Vbi(V)-V\right]\lambda_{\an(\cat)}}\right)^{2}}}
\label{eq:11}
\end{equation}
and $\Vbi(V)$ given by Eq.~(\ref{eq:9}). Alternatively, in terms of the inverse of the relative capacitance difference $\Delta C/\Cgeo$, Eq.~(\ref{eq:10}) can be rearranged as
\begin{equation}
\left[\frac{\Delta C(V)}{\Cgeo}\right]^{-1} = \frac{q}{2\eta(V)kT}\left[\Vbi(V)-V-\frac{2\eta(V)kT}{q}\right]
\label{eq:12}
\end{equation}
where $\Delta C(V) = C(V)-\Cgeo$ is the excess capacitance induced by injected charge carriers inside the active layer. The above analysis is valid for voltages that are smaller than few $kT/q$ below $\Vbi-2\eta kT/q$.

\begin{figure*}[t]
\centering
\includegraphics[width=0.86\textwidth]{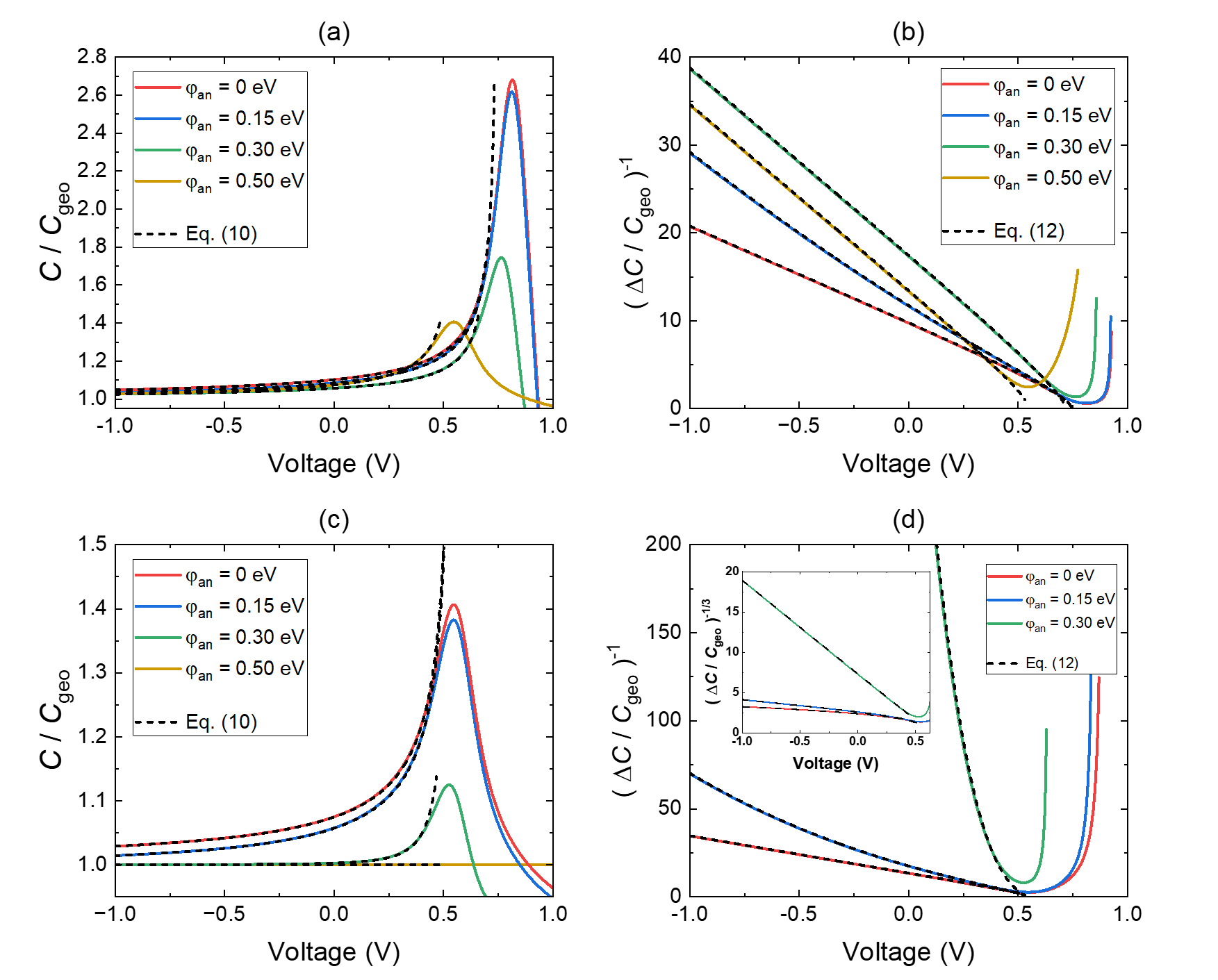}
\caption{Simulated $C/\Cgeo$ and $\left(\Delta C/\Cgeo\right)^{-1}$ as a function of voltage of a 100~nm-thick organic \mim{} diode is shown in (a) and (b), respectively, for the case of varying hole injection barrier at the anode contact, assuming the cathode contact remains ohmic for electron injection. (c) and (d) show an otherwise identical situation to (a) and (b), respectively, but for the case of a non-injecting cathode contact having a fixed electron injection barrier of 0.5~eV. Simulated curves (solid lines) are depicted by solid lines, while the analytical expressions Eq.~(\ref{eq:10}) and Eq.~(\ref{eq:12}) are indicated by the dashed lines. For comparison, corresponding $\left(\Delta C(V)/\Cgeo\right)^{-1/3}$ curves of the data in (c) and (d) are plotted in the inset of (d).}
\label{fig:2}
\end{figure*}

The analytical model [Eq.~(\ref{eq:10})] accurately reproduces the simulated voltage dependence of the capacitance for voltages below $\Vbi$, as shown in Fig.~\ref{fig:1}(a). Accordingly, $\Delta C(V)$ depends on both the dielectric properties of the active layer and the injecting contacts. Figure~\ref{fig:1}(c) shows the simulated \CV{} characteristics of a \mim{} diode with two ohmic contacts at different active layer thicknesses. The corresponding $\left(\Delta C/\Cgeo\right)^{-1}$ vs $V$ plot is shown in Fig.~\ref{fig:1}(d). The effect of a varying hole-injection barrier $\varphi_{\an}$ at the anode is demonstrated in Fig.~\ref{fig:2}, showing the simulated \CV{} characteristics and $\left(\Delta C/\Cgeo\right)^{-1}$ vs $V$ plots for the case of an electron-injecting and a non-injecting cathode contact. Comparison between numerical simulations (solid lines) and the analytical predictions (dashed lines) reveals an excellent agreement between the two, provided that the voltage is a few $kT/q$ below $\Vbi-2\eta kT/q$.

In the limit of ohmic contacts, $\eta$ becomes independent of voltage. If both contacts are ohmic, corresponding to $\lambda_{\an},\lambda_{\cat}\ll d$, then $\eta=2$. In this regime, the inverse of $\Delta C(V)$ is directly proportional to the electric field inside the active layer, $(\Delta C)^{-1}\propto\left[\Vbi(V)-V\right]$. Note that $\Vbi$ decreases with increasing $d$ as per Eq.~(\ref{eq:9}). Consequently, the \CV{} and $(\Delta C)^{-1}$ vs $V$ curves of the thicker devices in Fig.~\ref{fig:1}(c) and Fig.~\ref{fig:1}(d), respectively, are shifted to smaller voltages. As the injection barrier is gradually increased at one of the contacts, as shown Fig.~\ref{fig:2}(a) and (b), $\eta$ becomes voltage dependent and takes a value between $\eta=1$ and $\eta=2$. Eventually, as the injection barrier is large enough, corresponding to the case of a hole-only or electron-only diode with one ohmic and one non-injecting contact, $\eta$ again becomes independent of the applied voltage as $\eta\rightarrow 1$.

Finally, when both contacts become non-ohmic, a strong deviation from the ohmic contact regime is obtained. This is seen in Fig.~\ref{fig:2}(c) and (d), where a transition from the ohmic contact regime to the weakly-injecting contact regime is obtained as $\varphi_{\an}$ is increased. In the limit of two weakly-injecting contacts ($\lambda_{\an},\lambda_{\cat}\gg d$), Eq.~(\ref{eq:10}) reads
\begin{equation}
C(V) \approx \Cgeo + \frac{2(kT)^{2}d}{q\left[\Vbiz-V\right]^{3}}\left(p_{\an}+n_{\cat}\right)
\label{eq:13}
\end{equation}
with $\Delta C(V)$ instead obeying a $(\Delta C)^{-1/3}\propto\left[\Vbiz-V\right]$ type dependence; see inset of Fig.~\ref{fig:2}(d). In this limit, $\Delta C\ll\Cgeo$ generally applies, with $\Delta C$ being directly proportional to the carrier densities at the contacts, approaching $C\rightarrow\Cgeo$ as the injection barrier is increased. Note that the energy-level bending is negligible in this regime, with the carrier profiles being purely exponential.

The origin of the voltage dependence of the capacitance, seen in Fig.~\ref{fig:1} and \ref{fig:2}, can be traced back to the voltage modulation of the energy-level bending induced by the accumulation of injected carriers at the contacts. These carrier-accumulation regions are highly conductive, acting as virtual extensions of the contacts, effectively decreasing the thickness of the active layer. In accordance with Eq.~(\ref{eq:10}), the effective width of the hole (electron) accumulation region at the anode (cathode) contact can be written as $\Delta w_{\an(\cat)} = \frac{2kTd}{q\left[\Vbi-V\right]}\eta_{\an(\cat)}$. The accumulation regions become wider with increasing voltage and thickness, resulting in an increased capacitance. Larger $V$ and $d$ both translate into smaller magnitudes of the opposing electric field, allowing injected carriers to diffuse further into the active layer. Conversely, at high reverse bias, the strong electric field restricts the diffusion of injected carriers, shrinking the accumulation regions and, ultimately, leading to a saturation of the capacitance to $\Cgeo$ with increasing reverse-bias voltage.

The above analytical framework can be applied to probe the built-in potential within the device. It should be stressed, however, that both $\Vbi(V)$ and $\eta(V)$ have a non-negligible dependence on the voltage. To extract the (effective) built-in potential at $V=0$, we therefore expand Eq.~(\ref{eq:12}) to the first order to obtain a linear approximation for $\left[\Delta C(V)/\Cgeo\right]^{-1}$ in this voltage region (see Supplemental Material). Provided that at least one of the contacts is ohmic, Eq.~(\ref{eq:12}) can be approximated for voltages near $V=0$ by
\begin{equation}
\left[\frac{\Delta C(V)}{\Cgeo}\right]^{-1} \approx \frac{qC(0)}{2\eta kT\Cgeo}\left\{\Vbi-\frac{2\eta kT}{q}\left[1+\frac{\Cgeo}{C(0)}\right]-f_{S}V\right\}
\label{eq:14}
\end{equation}
with the notation $\Vbi = \Vbi(0)$ and $\eta = \eta(0)$, and where
\begin{equation}
f_{S} = \eta^{2}-6\left[\eta+\frac{1}{\eta}\right]+12
\label{eq:15}
\end{equation}
Note that Eq.~(\ref{eq:14}) takes the form $\left[\Delta C/\Cgeo\right]^{-1} = S\times\left(V^{*}-V\right)$, where $S$ is the slope and $V^{*}$ the $V$-intercept. Hence, from a linear fit of $\left[\Delta C(V)/\Cgeo\right]^{-1}$ near $V=0$, $\eta$ and $\Vbi$ can be deduced from the corresponding slope and intercept. Specifically, in the limiting case when one contact is ohmic and the other contact is either ohmic ($\eta=2$) or completely non-injecting ($\eta=1$), where $f_{S}\rightarrow 1$, we obtain $S = qC(0)/\left(2\eta kT\Cgeo\right)$. In this limit, the value of $\Vbi$ can be directly calculated from the estimated $S$ and $V^{*}$ using $\Vbi = V^{*}+S^{-1}\times\left[1+C(0)/\Cgeo\right]$.

\begin{figure*}[t]
\centering
\includegraphics[width=0.86\textwidth]{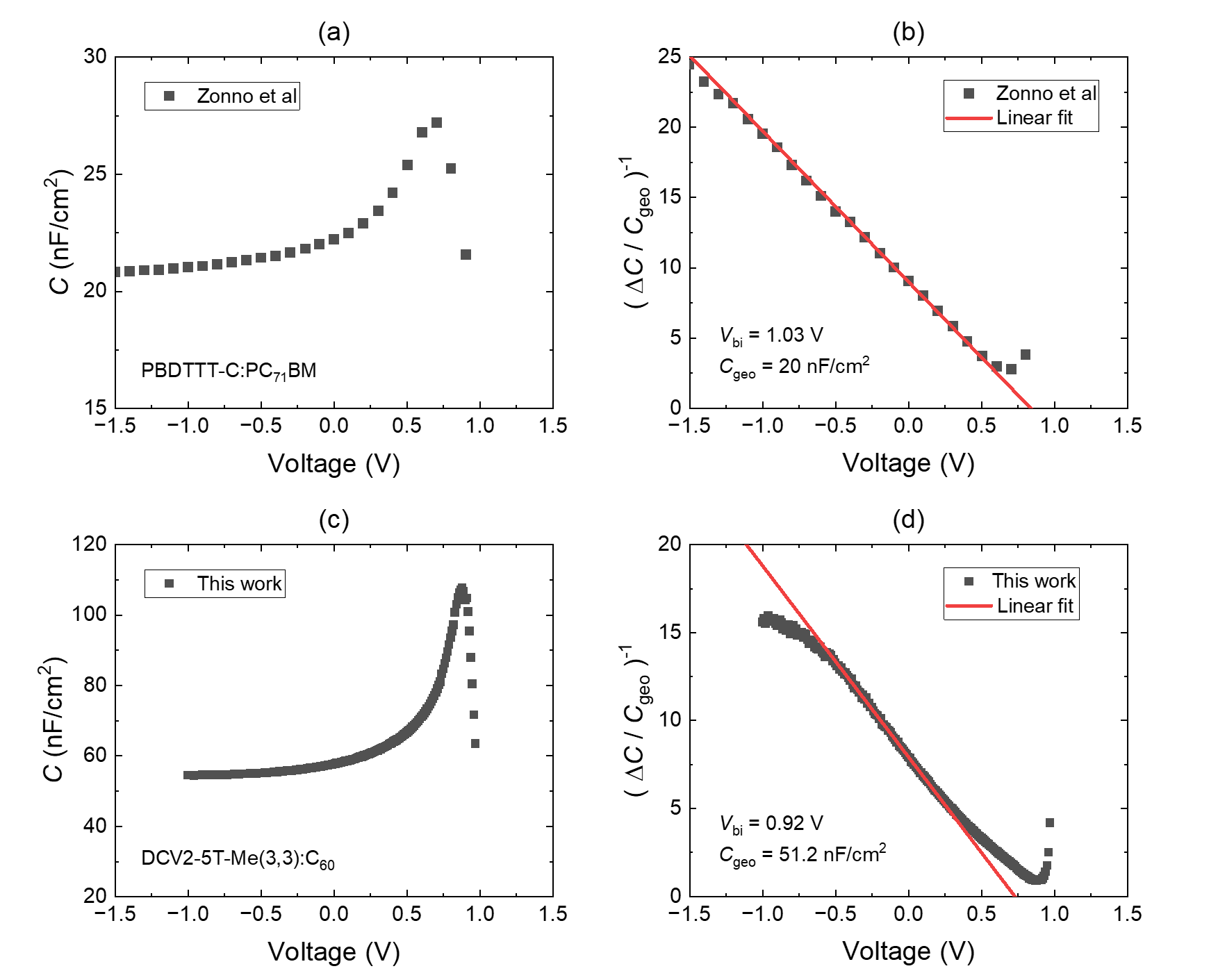}
\caption{Experimental \CV{} characteristics of organic bulk heterojunction solar cell device based on PBDTTT-C:PC$_{71}$BM, reported by Zonno \emph{et al.}~\cite{Zonno2017}, are shown in (a). The respective $\left(\Delta C/\Cgeo\right)^{-1}$ versus $V$ plot is shown in (b), assuming $\Cgeo = 20$~nF/cm$^{2}$. In (c), the measured \CV{} characteristics of a DCV$_{2}$-5T-Me(3,3):C$_{60}$ device are shown. The corresponding $\left(\Delta C/\Cgeo\right)^{-1}$ versus $V$ plot is shown in (d), with $\Cgeo = 51.2$~nF/cm$^{2}$. The solid lines indicate linear fits based on Eq.~(\ref{eq:14}) allowing for the $\Vbi$ at $V=0$~V to be extracted from the intercepts.}
\label{fig:3}
\end{figure*}

To substantiate the theoretical framework, we tested the method on organic bulk-heterojunction solar cells with undoped active layers. Figure~\ref{fig:3}(a) shows dark \CV{} characteristics of an ITO/ZnO/PBDTTT-C:PC$_{71}$BM(170~nm)/MoO$_{3}$/Ag device, reported by Zonno \emph{et al.}~\cite{Zonno2017} The corresponding $\left(\Delta C/\Cgeo\right)^{-1}$ versus $V$ plot is shown in Fig.~\ref{fig:3}(b) with an estimated $\Cgeo$ of 20~nF/cm$^{2}$. For comparison, the \CV{} characteristics of an 80~nm-thick DCV$_{2}$-5T-Me(3,3):C$_{60}$ device, measured in this work, is provided in Fig.~\ref{fig:3}(c). To ensure ohmic contacts, highly doped electrode interlayers were used for the DCV$_{2}$-5T-Me(3,3):C$_{60}$ device. Furthermore, a frequency of 100~Hz was used for the \CV{} measurements to ensure that the capacitance was not limited by charge carrier transport. For details on processing conditions \CV{} measurement setup, see Ref.~\cite{Wolansky2024} and Ref.~\cite{Kumar2026}, respectively. The $\left(\Delta C/\Cgeo\right)^{-1}$ versus $V$ plot for DCV$_{2}$-5T-Me(3,3):C$_{60}$ is shown in Fig.~\ref{fig:3}(d), with $\Cgeo = 51.2$~nF/cm$^{2}$. The experimental $\left(\Delta C/\Cgeo\right)^{-1}$ versus $V$ plots for both devices show a linear behaviour near $V=0$. The corresponding slopes suggest that $\eta=2$ for both cases (assuming $T=300$~K), consistent with bipolar \mim{} diodes with two ohmic injecting contacts. Subsequently, from the intercept, built-in potentials of $\Vbi = 1.03$~V for the PBDTTT-C:PC$_{71}$BM device and $\Vbi = 0.92$~V for the DCV$_{2}$-5T-Me(3,3):C$_{60}$ device were extracted at $V=0$. The obtained value for $\Vbi$ aligns well with the open-circuit voltage for the latter system~\cite{Wolansky2024,Kaienburg2022}.

In conclusion, the effect of injected carriers on the capacitance of thin-film \mim{} diode devices, based on undoped semiconductor active layers, has been clarified. After accounting for both the electrode charge and the space charge induced by the injected charge carriers within the active layer, an analytical description of the \CV{} characteristics is derived. Based on the analytical framework, a method to extract the built-in potential from \CV{} measurements in thin-film diode devices with intrinsic or lightly doped active layers (including organic solar cells and diodes) is presented. Additionally, the method can be used to identify and quantify ohmic contacts in these diodes. This method provides a complement to Mott-Schottky analysis that is only applicable to sufficiently doped active layers.

\begin{acknowledgments}
O.J.S. acknowledges funding from the Research Council of Finland through Project No.~357196. M.N. and J.B. acknowledge partial funding from the European Union's Horizon 2020 research and innovation programme, grant number 101008701 (EMERGE).
\end{acknowledgments}

\bibliography{refs}

@book{Forrest2000,
  author    = {Forrest, S. R.},
  title     = {Organic Electronics: Foundations to Applications},
  publisher = {Oxford University Press},
  address   = {New York},
  year      = {2000}
}

@book{Sze1981,
  author    = {Sze, S. M.},
  title     = {Physics of Semiconductor Devices},
  publisher = {Wiley and Sons},
  address   = {New York},
  year      = {1981}
}

@article{Neukom2018,
  author  = {Neukom, M. and Z\"ufle, S. and Jenatsch, S. and Ruhstaller, B.},
  title   = {Optoelectronic characterization of third-generation solar cells},
  journal = {Sci. Technol. Adv. Mater.},
  volume  = {19},
  pages   = {291--316},
  year    = {2018}
}

@article{Mingebach2011,
  author  = {Mingebach, M. and Deibel, C. and Dyakonov, V.},
  title   = {Built-in potential and validity of the {Mott-Schottky} analysis in organic bulk heterojunction solar cells},
  journal = {Phys. Rev. B},
  volume  = {84},
  pages   = {153201},
  year    = {2011}
}

@article{Kirchartz2012,
  author  = {Kirchartz, T. and Gong, W. and Hawks, S. A. and Agostinelli, T. and MacKenzie, R. C. I. and Yang, Y. and Nelson, J.},
  title   = {Sensitivity of the {Mott--Schottky} analysis in organic solar cells},
  journal = {J. Phys. Chem. C},
  volume  = {116},
  pages   = {7672},
  year    = {2012}
}

@article{Nigam2013,
  author  = {Nigam, A. and Premaratne, M. and Nair, P. R.},
  title   = {On the validity of unintentional doping densities extracted using {Mott--Schottky} analysis for thin film organic devices},
  journal = {Org. Electron.},
  volume  = {14},
  pages   = {2902--2907},
  year    = {2013}
}

@article{Kirchartz2015,
  author  = {Kirchartz, T. and Bisquert, J. and Mora-Sero, I. and Garcia-Belmonte, G.},
  title   = {Classification of solar cells according to mechanisms of charge separation and charge collection},
  journal = {Phys. Chem. Chem. Phys.},
  volume  = {17},
  pages   = {4007},
  year    = {2015}
}

@article{Dahlstrom2019,
  author  = {Dahlstr\"om, S. and Sandberg, O. J. and Nyman, M. and \"Osterbacka, R.},
  title   = {Determination of charge-carrier mobility and built-in potential in thin-film organic {M-I-M} diodes from extraction-current transients},
  journal = {Phys. Rev. Applied},
  volume  = {10},
  pages   = {054019},
  year    = {2019}
}

@article{Shao1961,
  author  = {Shao, J. and Wright, G. T.},
  title   = {Characteristics of the space-charge-limited dielectric diode at very high frequencies},
  journal = {Solid-State Electron.},
  volume  = {3},
  pages   = {291},
  year    = {1961}
}

@article{Shrotriya2005,
  author  = {Shrotriya, V. and Yang, Y.},
  title   = {Capacitance--voltage characterization of polymer light-emitting diodes},
  journal = {J. Appl. Phys.},
  volume  = {97},
  pages   = {054504},
  year    = {2005}
}

@article{vanMensfoort2008a,
  author  = {van Mensfoort, S. L. M. and Coehoorn, R.},
  title   = {Determination of injection barriers in organic semiconductor devices from capacitance measurements},
  journal = {Phys. Rev. Lett.},
  volume  = {100},
  pages   = {086802},
  year    = {2008}
}

@article{Kiermasch2018,
  author  = {Kiermasch, D. and Baumann, A. and Fischer, M. and Dyakonov, V. and Tvingstedt, K.},
  title   = {Revisiting lifetimes from transient electrical characterization of thin film solar cells; a capacitive concern evaluated for silicon, organic and perovskite devices},
  journal = {Energy Environ. Sci.},
  volume  = {11},
  pages   = {629--640},
  year    = {2018}
}

@article{Zeiske2022,
  author  = {Zeiske, S. and Sandberg, O. J. and Kurpiers, J. and Shoaee, S. and Meredith, P. and Armin, A.},
  title   = {Probing charge generation efficiency in thin-film solar cells by integral-mode transient charge extraction},
  journal = {ACS Photonics},
  volume  = {9},
  pages   = {1188--1195},
  year    = {2022}
}

@article{Bisquert2003,
  author  = {Bisquert, J.},
  title   = {Chemical capacitance of nanostructured semiconductors: its origin and significance for nanocomposite solar cells},
  journal = {Phys. Chem. Chem. Phys.},
  volume  = {5},
  pages   = {5360},
  year    = {2003}
}

@article{Tripathi2013,
  author  = {Tripathi, D. C. and Mohapatra, Y. N.},
  title   = {Diffusive capacitance in space charge limited organic diodes: analysis of peak in capacitance-voltage characteristics},
  journal = {Appl. Phys. Lett.},
  volume  = {102},
  pages   = {253303},
  year    = {2013}
}

@article{Nandal2018,
  author  = {Nandal, V. and Nair, P. R.},
  title   = {Anomalous scaling exponents in the capacitance-voltage characteristics of perovskite thin film devices},
  journal = {J. Phys. Chem. C},
  volume  = {122},
  pages   = {27935--27940},
  year    = {2018}
}

@article{vanMensfoort2008b,
  author  = {van Mensfoort, S. L. M. and Coehoorn, R.},
  title   = {Effect of {Gaussian} disorder on the voltage dependence of the current density in sandwich-type devices based on organic semiconductors},
  journal = {Phys. Rev. B},
  volume  = {78},
  pages   = {085207},
  year    = {2008}
}

@article{Lange2011,
  author  = {Lange, I. and Blakesley, J. C. and Frisch, J. and Vollmer, A. and Koch, N. and Neher, D.},
  title   = {Band bending in conjugated polymer layers},
  journal = {Phys. Rev. Lett.},
  volume  = {106},
  pages   = {216402},
  year    = {2011}
}

@article{deBruyn2013,
  author  = {de Bruyn, P. and van Rest, A. H. P. and Wetzelaer, G. A. H. and de Leeuw, D. M. and Blom, P. W. M.},
  title   = {Diffusion-limited current in organic metal-insulator-metal diodes},
  journal = {Phys. Rev. Lett.},
  volume  = {111},
  pages   = {186801},
  year    = {2013}
}

@article{Sandberg2024,
  author  = {Sandberg, O. J. and Armin, A.},
  title   = {Diode equation for sandwich-type thin-film photovoltaic devices limited by bimolecular recombination},
  journal = {PRX Energy},
  volume  = {3},
  pages   = {023008},
  year    = {2024}
}

@article{Hawks2015,
  author  = {Hawks, S. A. and Finck, B. Y. and Schwartz, B. J.},
  title   = {Theory of current transients in planar semiconductor devices: insights and applications to organic solar cells},
  journal = {Phys. Rev. Applied},
  volume  = {3},
  pages   = {044014},
  year    = {2015}
}

@article{Zonno2019,
  author  = {Zonno, I. and Zayani, H. and Grzeslo, M. and Krogmeier, B. and Kirchartz, T.},
  title   = {Extracting recombination parameters from impedance measurements on organic solar cells},
  journal = {Phys. Rev. Applied},
  volume  = {11},
  pages   = {054024},
  year    = {2019}
}

@article{Sandberg2019,
  author  = {Sandberg, O. J. and Tvingstedt, K. and Meredith, P. and Armin, A.},
  title   = {Theoretical perspective on transient photovoltage and charge extraction techniques},
  journal = {J. Phys. Chem. C},
  volume  = {123},
  pages   = {14261--14271},
  year    = {2019}
}

@misc{SupplementalMaterial,
  note = {See Supplemental Material for additional derivations and approximations}
}

@article{Hartnagel2023,
  author  = {Hartnagel, P. and Ravishankar, S. and Klingebiel, B. and Thimm, O. and Kirchartz, T.},
  title   = {Comparing methods of characterizing energetic disorder in organic solar cells},
  journal = {Adv. Energy Mater.},
  volume  = {13},
  pages   = {2300329},
  year    = {2023}
}

@article{deLevie1972a,
  author  = {de Levie, R. and Moreira, H.},
  title   = {Transport of ions of one kind through thin membranes: {I}. General and equilibrium considerations},
  journal = {J. Membr. Biol.},
  volume  = {9},
  pages   = {241},
  year    = {1972}
}

@article{deLevie1972b,
  author  = {de Levie, R. and Seidah, N. G. and Moreira, H.},
  title   = {Transport of ions of one kind through thin membranes: {II}. Nonequilibrium steady-state behavior},
  journal = {J. Membr. Biol.},
  volume  = {10},
  pages   = {171},
  year    = {1972}
}

@article{Cheyns2008,
  author  = {Cheyns, D. and Poortmans, J. and Heremans, P. and Deibel, C. and Verlaak, S. and Rand, B. P. and Genoe, J.},
  title   = {Analytical model for the open-circuit voltage and its associated resistance in organic planar heterojunction solar cells},
  journal = {Phys. Rev. B},
  volume  = {77},
  pages   = {165332},
  year    = {2008}
}

@article{Zonno2017,
  author  = {Zonno, I. and Martinez-Otero, A. and Hebig, J.-C. and Kirchartz, T.},
  title   = {Understanding {Mott-Schottky} measurements under illumination in organic bulk heterojunction solar cells},
  journal = {Phys. Rev. Applied},
  volume  = {7},
  pages   = {034018},
  year    = {2017}
}

@article{Wolansky2024,
  author  = {Wolansky, J. and Hoffmann, C. and Panhans, M. and Winkler, L. C. and Talnack, F. and Hutsch, S. and Zhang, H. and Kirch, A. and Yallum, K. M. and Friedrich, H. and Kublitski, J. and Gao, F. and Spoltore, D. and Mannsfeld, S. C. B. and Ortmann, F. and Banerji, N. and Leo, K. and Benduhn, J.},
  title   = {Sensitive self-driven single-component organic photodetector based on vapor-deposited small molecules},
  journal = {Adv. Mater.},
  volume  = {36},
  pages   = {2402834},
  year    = {2024}
}

@article{Kumar2026,
  author  = {Kumar, M. and Ding, C. and Yang, J. and Sandberg, O. J. and Nyman, M. and Ma, C.-Q. and \"Osterbacka, R.},
  title   = {Effect of thermal stress on the ion density and mobility distribution in perovskite solar cells},
  journal = {J. Phys. Chem. Lett.},
  volume  = {17},
  pages   = {1527--1533},
  year    = {2026}
}

@article{Kaienburg2022,
  author  = {Kaienburg, P. and Jungbluth, A. and Habib, I. and Vajjala Kesava, S. and Nyman, M. and Riede, M. K.},
  title   = {Assessing the photovoltaic quality of vacuum-thermal evaporated organic semiconductor blends},
  journal = {Adv. Mater.},
  volume  = {34},
  pages   = {2107584},
  year    = {2022}
}

\end{document}